\documentclass[11pt, a4paper]{article}

\usepackage{graphicx}
\usepackage{amsmath, amssymb, amsthm}
\usepackage[margin=2.5cm]{geometry}
\usepackage{hyperref}
\usepackage{booktabs}
\usepackage{caption}
\usepackage{subcaption}
\usepackage{authblk}
\usepackage[numbers,sort&compress]{natbib}
\usepackage{xcolor}
\usepackage{algorithm}
\usepackage{algpseudocode}
\usepackage{enumitem}

\hypersetup{
    colorlinks=true,
    linkcolor=blue,
    filecolor=magenta,      
    urlcolor=cyan,
    citecolor=blue
}

\theoremstyle{definition}
\newtheorem{definition}{Definition}[section]
\theoremstyle{plain}
\newtheorem{proposition}{Proposition}[section]

\newtheorem{remark}{Remark}[section]

\title{\textbf{Localizing Preference Aggregation Conflicts:\\ A Graph-Theoretic Approach Using Sheaves}}

\author[1]{Karen Sargsyan\thanks{Corresponding author: karen.sarkisyan@gmail.com}}
\affil[1]{Institute of Chemistry, Academia Sinica, Taipei, Taiwan}

\begin{document}

\maketitle

\begin{abstract}
We introduce a graph-theoretic framework based on discrete sheaves to diagnose and localize inconsistencies in preference aggregation and, more broadly, in the fusion of partial rankings supplied by many overlapping sources. Unlike linearization methods such as HodgeRank, which embed comparisons into a numerical flow, this approach stays purely ordinal and locates conflict in the interaction structure via the Obstruction Locus, identifying which voter pairs fail to cohere. We formalize the Incompatibility Index to quantify these local conflicts and examine their behavior under stochastic variations using the Mallows model. We further develop a sheaf-theoretic pushforward operation to model voter merging, implemented via a polynomial-time constraint digraph algorithm. We demonstrate that graph quotients transform distributed edge conflicts into local impossibilities (empty stalks), showing topologically how aggregation paradoxes can persist across scales.
\end{abstract}

\section{Introduction}

The challenge of aggregating individual preferences into a collective decision is central to social choice theory. Since the seminal work of Condorcet \cite{Condorcet1785} and the foundational impossibility theorem of Arrow \cite{Arrow1951}, it has been understood that ordinal preference aggregation is inherently susceptible to paradoxes and inconsistencies. These paradoxes typically manifest as cycles (e.g., A preferred over B, B over C, and C over A), obstructing the formation of a coherent global order.

In many contemporary settings the task is a different one. The goal is not to settle a contest among voters who may legitimately disagree, but to \emph{fuse} ordinal data spread across many partially overlapping sources into one coherent ranking. Several search engines or recommender systems each rank only the items they have indexed; a program committee gives each reviewer only part of the submission pile; a panel of raters each orders the subset of items it has seen. Two features are common to these examples. Every source observes only a \emph{subset} of the alternatives, so partial visibility is built into the data and not an extra assumption. And whenever two sources share alternatives, they have to agree on the shared ones for a single global ranking to exist. A disagreement on such an overlap is no longer dissent to be out-voted; it is an \emph{inconsistency} that blocks fusion. What one then wants to know is \emph{where} the data fail to cohere, so that the conflict can be located and repaired. That is the problem we study, and the Condorcet cycle turns out to be its smallest nontrivial instance.

The inherent complexity of this problem has spurred the application of diverse mathematical frameworks, increasingly drawing from topology and geometry to understand the structure of preference spaces and the nature of these obstructions \cite{Chichilnisky1980, Baigent1987}. One prominent approach involves adapting Hodge theory to combinatorial settings, often referred to as ``HodgeRank'' \cite{Jiang2011}. This method linearizes the preference data, mapping ordinal comparisons to numerical flows on a graph. While powerful, linearization can obscure the discrete, combinatorial nature of ordinal preferences and may prioritize global topological features over the specific localization of inconsistencies within the profile.

In parallel, sheaf theory has emerged as a powerful mathematical framework for studying local-to-global problems \cite{Bredon1997, Kashiwara2006, Rosiak2022}. A sheaf provides a formal mechanism to track data assigned to different parts of a space and enforce compatibility on their overlaps. Its application has expanded significantly in applied topology, addressing problems in data fusion, networks, and dynamics \cite{Ghrist2014, Curry2014, Hansen2020}.

This study develops a discrete sheaf-theoretic framework for the fusion problem that works directly with the ordinal data, without linearizing it. The sources are the vertices of a graph $G=(V,E)$; each vertex carries a single local total order over the alternatives it observes, and two vertices share an edge when they observe a common alternative and so are required to agree there. The candidate edges are grounded in the data: two sources can be asked to agree only where they actually share alternatives, so the maximal interaction graph, one edge per overlapping pair, is determined by the visibility sets, and any graph we use is a subgraph of it. Which overlaps to enforce remains a modeling choice (Section~\ref{sec:limitations}), but a constrained one. (We call a source a \emph{voter} where the social-choice reading is natural; the terms are interchangeable.) We introduce the concept of the \textit{Obstruction Locus} ($\Omega$) to quantify and localize inconsistencies, and demonstrate through computational experiments that this framework reveals interesting phenomena in the structure of preference profiles. Our main contributions are:

\begin{enumerate}
    \item The definition and implementation of the Discrete Order Sheaf for preference profiles.
    \item The definition of the Obstruction Locus ($\Omega_1$) and the Incompatibility Index ($|\Omega_1|$), with clear explanation of what this quantity measures.
    \item Experimental evidence showing how $|\Omega_1|$ varies with profile changes, using both a stochastic interpolation model and illustrative deterministic examples.
    \item Implementation of the sheaf-theoretic pushforward ($\pi_*$), revealing that merging conflicts transforms edge obstructions into local impossibilities (empty stalks).
\end{enumerate}

\paragraph{What the sheaf framework provides (and what it does not).}
Let us state the scope plainly. The Incompatibility Index $|\Omega_1|$ is not a cohomological invariant. It is the size of the support of the coboundary $\delta\sigma$, a count of the edges on which the data fail to cohere, and for sheaves of total orders the higher obstructions vanish, so the full machinery of sheaf cohomology never comes into play. The framework still earns two things that a bare inconsistency score does not. One is localization: a single global number such as the harmonic norm of HodgeRank reports only that some obstruction exists, whereas the coboundary support says which interactions are responsible for it. The other is functoriality: the pushforward $\pi_*$ along a quotient map follows, in polynomial time, what becomes of these conflicts when sources are merged, as distributed edge obstructions turn into empty stalks. These are the two properties we actually rely on. In particular, $|\Omega_1|=0$ certifies pairwise coherence but \emph{not} global fusability: sources can agree on every overlap and still admit no common ranking (Appendix~\ref{appendix:example2}). Closing that gap is exactly what the pushforward is for---it is the certificate for global solvability, while $|\Omega_1|$ is the localizer of pairwise conflict.

\section{Related Work}

The intersection of topology, geometry, and social choice theory is a rich and growing area of research. Our work builds upon several distinct traditions.

\subsection{Topological Social Choice}

The application of topological methods to social choice was pioneered by Chichilnisky \cite{Chichilnisky1980}, who demonstrated that the topological properties of the space of preferences impose constraints on the existence of continuous aggregation rules satisfying certain axioms (like anonymity and unanimity). Chichilnisky and Heal \cite{ChichilniskyHeal1983} subsequently established that contractibility of the preference space is necessary and sufficient for the existence of such a rule. Baigent \cite{Baigent1987} explored preference proximity and its implications for social choice rules, and Weinberger \cite{Weinberger2004} gave an equivariant-topology treatment of the model, extending the Chichilnisky--Heal framework. Our work builds on this tradition but differs by employing sheaf-theoretic localization specifically to identify which edges in the interaction graph cause inconsistencies.

\subsection{Applied Sheaf Theory}

Sheaf theory, originally developed in algebraic geometry \cite{Bredon1997, Kashiwara2006}, has recently found significant applications in applied topology. Ghrist emphasized the utility of sheaves for understanding distributed systems and sensor networks \cite{Ghrist2014}. Curry provided foundational work on sheaves and cosheaves in applied settings, exploring their connection to persistence and data fusion \cite{Curry2014}. Robinson investigated assignments to sheaves of pseudometric spaces, relevant for modeling data with inherent geometric structure \cite{Robinson2020}. More recently, Hansen and Ghrist explored opinion dynamics using discourse sheaves, modeling the evolution of beliefs over networks \cite{Hansen2020}. Our framework extends these ideas to the domain of ordinal preference aggregation.

\subsection{Hodge Theory and Ranking}

The adaptation of Hodge theory to discrete settings, particularly for ranking from pairwise comparisons, was formalized by Jiang et al. with HodgeRank \cite{Jiang2011}. This approach linearizes preference data into flows on a graph and applies the Helmholtz--Hodge decomposition, splitting a flow into a gradient component (the globally consistent ranking), a curl component (local, e.g.\ triangular, intransitivities), and a harmonic component (globally cyclic inconsistencies with no local witness). Inconsistency is then quantified through the $L_2$ norms of the curl and harmonic parts, the harmonic part capturing specifically the global cyclic obstructions. While effective, Hodge-theoretic methods rely on linearization. As discussed in Section~\ref{sec:results_localization}, our discrete approach localizes conflict differently, among voter interactions rather than among alternatives, and stays purely ordinal.

\subsection{Judgment Aggregation and Logical Constraints}

Beyond preference aggregation, the broader field of judgment aggregation studies how to combine individual judgments on logically connected propositions \cite{ListPettit2002, ListPuppe2009}. The doctrinal paradox and related impossibility results \cite{ListPettit2002} demonstrate that logical constraints between propositions can obstruct consistent aggregation, analogous to how preference cycles obstruct ordinal aggregation. Our sheaf-theoretic framework naturally accommodates such logical constraints through the restriction maps, and the Obstruction Locus can be interpreted as identifying where logical consistency fails across the network.

\subsection{Algebraic Approaches to Arrow's Theorem}

Arrow's theorem has been analyzed through various algebraic lenses\cite{Fishburn1970, KirmanSondermann1972, Lauwers1995}. The ultrafilter proof of Arrow's theorem \cite{KirmanSondermann1972} characterizes dictatorships as ultrafilters on the set of voters, providing an elegant algebraic perspective. More recent work by Lauwers and Van Liedekerke \cite{Lauwers1995} recasts Arrovian aggregation in the language of model theory and ultraproducts. While our approach differs---we focus on localization of conflicts rather than characterization of aggregation rules---these algebraic perspectives inform potential future connections.

\subsection{Probabilistic Models of Preferences}

The Mallows model \cite{Mallows1957} provides a probability distribution over permutations centered on a reference ordering, with a dispersion parameter controlling concentration. This model has been proposed in machine learning and statistics for analyzing ranking data \cite{Lu2011}. We employ the Mallows model in our stochastic interpolation experiments to generate realistic preference distributions with controllable consensus levels.

\section{Framework and Methods}

We model a population of voters and their interactions as a graph $G=(V,E)$, where vertices $V$ represent voters and edges $E$ represent interactions (overlaps or shared comparisons) where consistency is required. The set of alternatives over which voters may have preferences is denoted by $\mathcal{A}$.

\subsection{The Discrete Order Sheaf}

\begin{definition}[Discrete Order Presheaf]
Let $F$ be a cellular presheaf on the graph $G=(V,E)$---that is, a presheaf on the \emph{face poset} of $G$, the partially ordered set whose elements are the vertices and edges of $G$, ordered by incidence ($v \le e$ whenever $v$ is an endpoint of $e$). Concretely, $F$ assigns data to each vertex and each edge, together with a restriction map for every incidence $v \le e$. For each vertex $v \in V$, let $\mathcal{A}_v \subseteq \mathcal{A}$ denote the subset of alternatives visible at $v$. We define:

\begin{enumerate}
    \item \textbf{Vertex stalks:} $F_v$ is the set of all total orders on $\mathcal{A}_v$.
    
    \item \textbf{Edge stalks:} For each edge $e = \{u,v\} \in E$, let $\mathcal{A}_e = \mathcal{A}_u \cap \mathcal{A}_v$ be the overlap of visible alternatives. Then $F_e$ is the set of all total orders on $\mathcal{A}_e$.
    
    \item \textbf{Restriction maps:} For each incidence $v \subseteq e$ (i.e., vertex $v$ incident to edge $e = \{u,v\}$), the restriction map $\rho^v_e: F_v \to F_e$ is given by:
    \[
    \rho^v_e(\sigma_v) = \sigma_v|_{\mathcal{A}_e}
    \]
    where $\sigma_v|_{\mathcal{A}_e}$ denotes the restriction of the total order $\sigma_v$ to the subset $\mathcal{A}_e \subseteq \mathcal{A}_v$. We denote by $\mathcal{T}(S)$ the set of all total orders on a finite set $S$.
\end{enumerate}
\end{definition}

\begin{remark}[$F$ is a cellular sheaf]
The data above, a set of orders on each cell (vertex or edge) together with a restriction map for each incidence $v \le e$, is precisely a \emph{cellular sheaf} on $G$ \cite{Curry2014}. There is no separation or gluing axiom to verify here: sheaves on the Alexandrov topology of a poset are the same as functors on that poset, so once the stalks and restriction maps are specified the sheaf is determined, with nothing further to check. The mathematical content lies not in whether $F$ is a sheaf but in describing its global sections, which is governed by the coboundary $\delta$ and the Obstruction Locus introduced next.
\end{remark}

\begin{definition}[Preference Profile as a 0-Cochain]
A preference profile $\sigma$ is an assignment of a specific local order $\sigma_v \in F_v$ to each vertex $v$. In the language of the \v{C}ech complex, $\sigma$ is a 0-cochain: $\sigma \in C^0(G, F) = \prod_{v \in V} F_v$.
\end{definition}

\subsection{The Obstruction Locus ($\Omega$)}
\label{sec:obstruction}

We define compatibility constraints using the coboundary operator $\delta$. For an edge $e=\{u,v\}$, the overlap is the intersection of visible alternatives $\mathcal{A}_{uv} = \mathcal{A}_u \cap \mathcal{A}_v$.

\begin{definition}[Obstruction Locus $\Omega_1$ and Incompatibility Index]
\label{def:omega}
The edge Obstruction Locus $\Omega_1(\sigma)$ is the set of edges where local orders are incompatible:
$$ \Omega_1(\sigma) = \{e=\{u,v\} \in E \mid \sigma_u|_{\mathcal{A}_{uv}} \neq \sigma_v|_{\mathcal{A}_{uv}} \} $$
The Incompatibility Index $|\Omega_1(\sigma)|$ is the cardinality of this set. When $|\mathcal{A}_{uv}| \leq 1$ the two restricted orders coincide trivially, so such edges are never obstructed.
\end{definition}

\begin{remark}[What $|\Omega_1|$ Measures]
\label{remark:what_omega_measures}
The Incompatibility Index $|\Omega_1(\sigma)|$ counts the number of edges where adjacent voters disagree on their shared alternatives. This is the support of the coboundary $\delta\sigma$, not a cohomological dimension. The advantage of this measure is localization: it identifies which specific edges cause inconsistency rather than merely detecting that inconsistency exists.
\end{remark}

On a graph there are no cells above dimension one, so the higher obstruction loci $\Omega_k$ ($k\ge 2$) are empty for dimensional reasons. Even on a higher-dimensional complex the agreement-on-overlaps that $\Omega_k$ records would vanish trivially: a triple overlap $\mathcal{A}_{uvw}$ is contained in each pairwise overlap, so agreement on edges already forces agreement there. This is a property of the \v{C}ech construction, not of total orders specifically---it holds for sheaves of partial or interval orders as well. We stress that this triviality of $\Omega_k$ does \emph{not} mean that pairwise-consistent total orders glue freely: they may still fail to assemble into a single global ranking, which is precisely the empty-stalk phenomenon the pushforward detects (Sections~\ref{sec:pushforward} and~\ref{sec:why_sheaf}). Capturing nontrivial \emph{higher} obstructions would require a finer construction, which we leave to future work. Algorithmic details for computing $\Omega_1$ are provided in Appendix \ref{appendix:algorithms}.

\subsection{Global Consistency (H$^0$)}

\begin{proposition}[Global Consistency of a Profile]
A profile $\sigma$ is a global section, $\sigma \in H^0(G,F)$, if and only if its Obstruction Locus is empty: $\Omega(\sigma) = \emptyset$. (Since $\sigma$ already selects an order at each vertex, the stalks $F_v$ are automatically nonempty here; nonemptiness becomes a substantive hypothesis only for the pushforward sheaf, where a stalk can be empty.)
\end{proposition}

\begin{remark}
For the base Discrete Order Sheaf a global section always exists (any total order on $\mathcal{A}$ restricts to a compatible profile), so $H^0(G,F) \neq \emptyset$ unconditionally. The substantive question is therefore not whether \emph{some} section exists but whether the \emph{observed} profile $\sigma$ is one, which is exactly the condition $\Omega(\sigma)=\emptyset$. The nonempty-stalk hypothesis is automatic here, since total orders exist on any finite nonempty $\mathcal{A}_v$. It becomes substantive only for the pushforward sheaf $\pi_* F^\sigma$, where a stalk can be empty due to contradictory constraints; there the existence of any global section can genuinely fail.
\end{remark}

\subsection{The Pushforward (\texorpdfstring{$\pi_*$}{pi*})}
\label{sec:pushforward}

Given a graph quotient map $\pi: G \to G'$ (e.g., a vertex merge), the pushforward sheaf $\pi_* F$ on $G'$ is defined as follows.

\begin{definition}[Pushforward Sheaf $\pi_* F^\sigma$]
Let $\pi: G \to G'$ be a graph quotient map (e.g., vertex contraction). Given a preference profile $\sigma$ on $G$, we define the pushforward sheaf restricted to $\sigma$, denoted $\pi_* F^\sigma$, on $G'$ as follows:

\begin{enumerate}
    \item \textbf{Vertex Stalks:} For a vertex $v' \in V(G')$, let $P = \pi^{-1}(v') \subseteq V(G)$ be its preimage. The set of visible alternatives at $v'$ is $\mathcal{A}_{v'} = \bigcup_{v \in P} \mathcal{A}_v$. The stalk is the set of total orders on $\mathcal{A}_{v'}$ compatible with all original voters in the preimage:
    \[
    (\pi_* F^\sigma)_{v'} = \{ \tau \in \mathcal{T}(\mathcal{A}_{v'}) \mid \forall v \in P, \tau|_{\mathcal{A}_v} = \sigma_v \}
    \]
    where the restriction $\tau|_{\mathcal{A}_v}$ denotes the induced order on $\mathcal{A}_v \subseteq \mathcal{A}_{v'}$.
    
    \item \textbf{Edge Stalks:} For an edge $e' = \{u', v'\} \in E(G')$, the stalk is the set of all total orders on the overlap $\mathcal{A}_{e'} = \mathcal{A}_{u'} \cap \mathcal{A}_{v'}$:
    \[
    (\pi_* F^\sigma)_{e'} = \mathcal{T}(\mathcal{A}_{e'}).
    \]
    
    \item \textbf{Restriction Maps:} For an incidence $v' \in e'$, the restriction map $r: (\pi_* F^\sigma)_{v'} \to (\pi_* F^\sigma)_{e'}$ is the standard restriction of a total order to a subset of alternatives.
\end{enumerate}

\noindent If the constraints in the preimage $P$ are contradictory (for instance, when they form a directed cycle in the constraint digraph introduced below), the vertex stalk $(\pi_* F^\sigma)_{v'}$ is the empty set $\emptyset$. The obstruction is then localized \emph{at the merged vertex}: the incident edge stalks remain full, so the edge Incompatibility Index can vanish even though no global section exists.

\begin{remark}[Convention for $\Omega_1$ with empty stalks]
\label{rem:omega_convention}
The Incompatibility Index of Definition~\ref{def:omega} compares the restrictions $\sigma_u|_{\mathcal{A}_{uv}}$ and $\sigma_v|_{\mathcal{A}_{uv}}$, which presupposes a chosen order at each endpoint. When a pushforward stalk is empty there is no order to restrict, so we adopt the convention that $\Omega_1$ on the quotient counts only edges \emph{both} of whose endpoint stalks are nonempty. An empty stalk therefore contributes no edge obstruction. This is the precise sense in which $|\Omega_1|$ can be $0$ on the quotient while $H^0=\emptyset$: the obstruction has left the edges entirely and lives in the empty stalk.
\end{remark}
\end{definition}

\begin{proposition}[Merges compose]
\label{prop:merges_compose}
Fix a profile $\sigma$ on $G$. For a set $S \subseteq V(G)$ of original vertices, write
\[
M(S) = \bigl\{\, \tau \in \mathcal{T}(\textstyle\bigcup_{v\in S} \mathcal{A}_v) \;\big|\; \tau|_{\mathcal{A}_v} = \sigma_v \text{ for all } v \in S \,\bigr\}
\]
for the order set obtained by merging the vertices in $S$. If a vertex $v''$ of an iterated quotient $\pi' \circ \pi : G \to G''$ has total preimage $S = (\pi'\circ\pi)^{-1}(v'')$, then $\bigl((\pi'\circ\pi)_* F^\sigma\bigr)_{v''} = M(S)$. Consequently the merged stalk---and in particular whether it is empty---depends only on which original vertices are identified, not on the order or grouping in which the merges are carried out.
\end{proposition}

\begin{proof}
By the definition of the pushforward, the stalk over $v''$ is the set of orders on $\bigcup_{v\in S}\mathcal{A}_v$ that restrict to $\sigma_v$ on each $\mathcal{A}_v$, $v \in S$, which is exactly $M(S)$. The defining constraints are the union of the pairwise precedence requirements contributed by the local orders $\{\sigma_v : v\in S\}$, a set determined by $S$ alone. Hence $M(S)$, and its emptiness, is independent of any intermediate grouping $\pi$.
\end{proof}

\subsubsection{Computational Method: Constraint Digraph Approach}

Computing the pushforward stalk efficiently requires determining which total orders on the merged alternative set $\mathcal{A}_{v'} = \bigcup_{v \in P} \mathcal{A}_v$ are compatible with all local orders in the preimage $P$. A naive approach would enumerate all $|\mathcal{A}_{v'}|!$ permutations, which is exponential and infeasible for large $|\mathcal{A}_{v'}|$. Instead, we employ a \emph{constraint digraph} approach (the digraph is a directed acyclic graph, or DAG, precisely when the constraints are satisfiable). Each local order in the preimage imposes pairwise ordering constraints. For a voter with order $a_1 > a_2 > \cdots > a_k$, we have constraints $a_i > a_{i+1}$ for all $i$. Compatibility requires that any global order respects all such constraints from all voters in the preimage.

\textbf{Algorithm Overview:}
\begin{enumerate}
    \item \textbf{Build constraint digraph $C$:} Create a directed graph with vertices $\mathcal{A}_{v'}$ and add edge $(a,b)$ for each pairwise constraint ``$a$ must precede $b$'' from any voter in the preimage.
    
    \item \textbf{Detect cycles:} If $C$ contains a directed cycle, the constraints are contradictory and the stalk is empty. This occurs when voters have conflicting preferences (e.g., $V_1$ requires $A > B$ while $V_2$ requires $B > A$).
    
    \item \textbf{Find compatible orders:} If $C$ is acyclic (a DAG), any topological ordering of $C$ yields a compatible total order. The set of all topological orderings constitutes the pushforward stalk.
\end{enumerate}

This approach reduces the complexity from $O(|\mathcal{A}|!)$ to $O(|\mathcal{A}|^2 \cdot |P|)$ for cycle detection, where $|P|$ is the size of the preimage. Building the digraph requires $O(|\mathcal{A}| \cdot |P|)$ time (adding edges for each voter's constraints), and cycle detection via depth-first search is $O(|\mathcal{A}| + |E_C|)$ where $|E_C| \leq |\mathcal{A}| \cdot |P|$ is the number of constraint edges.

\begin{remark}[Constraint Digraph as Intersection of Constraints]
The definition of the pushforward stalk $(\pi_* F^\sigma)_{v'}$ requires finding an order $\tau$ that satisfies multiple local conditions simultaneously. In set-theoretic terms, the stalk is the intersection of the solution sets of individual voters. The constraint digraph algorithm computes this intersection efficiently. A cycle in the digraph corresponds to an empty intersection ($\bigcap \text{valid orders} = \emptyset$), providing a computational certificate that the stalk is empty.
\end{remark}

\begin{remark}[Counting Compatible Orders]
\label{rem:counting}
While cycle detection is polynomial, \textit{counting} the compatible total orders is \#P-complete \cite{BrightwellWinkler1991}, since it amounts to counting the linear extensions of the constraint digraph. (Listing those orders is a separate matter: there can be exponentially many of them, so enumeration is unavoidably exponential regardless of how hard counting turns out to be.) For most practical preference profiles the constraint digraph is highly constrained, yielding few compatible orders. Our implementation returns one witness ordering for large $|\mathcal{A}|$, which suffices for existence proofs and empty stalk detection.
\end{remark}

The complete algorithm with complexity analysis appears in Appendix~\ref{appendix:algorithms}; a worked example demonstrating empty stalk detection is provided in Appendix~\ref{appendix:example}.

\subsection{Why not just compare edges directly?}
\label{sec:why_sheaf}

Having built the stalks, restriction maps, coboundary, and pushforward, it is fair to ask whether that apparatus buys anything over a direct edge comparison. Computing $\Omega_1$ comes down to looping over the edges and comparing two restricted orders, and for that one count on a fixed graph the sheaf vocabulary really is dispensable; we do not pretend otherwise. Its value shows up in the questions that an edge-by-edge view cannot even phrase.

The first such question is global solvability. What we usually want to know is whether the partial data assemble into one coherent ranking, that is, whether $H^0$ is nonempty. Pairwise comparison reports where two sources clash, but a profile can have no pairwise clashes and still admit no global section once stalks are allowed to be empty. Solvability is settled by the joint condition of an empty obstruction locus and nonempty stalks (the Global Consistency proposition above), not by edge agreement on its own.

The second reason matters more. Conflicts do not stay put when the graph is coarsened. Merge two sources under a quotient map $\pi$ and the edge that recorded their disagreement is gone, but the disagreement is not: the pushforward $\pi_* F$ moves it into the stalk of the merged vertex, which then empties out (Section~\ref{sec:pushforward}). An accounting scheme that only knows how to compare the two ends of an edge has nothing to say about an obstruction that now sits at a vertex. Following the same conflict as it changes form under coarsening is what the pushforward gives us: merges compose cleanly (Proposition~\ref{prop:merges_compose}), so the obstruction is tracked consistently across successive quotients no matter how the merging is staged. The gap is sharpest when there is no obstructed edge to begin with: Appendix~\ref{appendix:example2} exhibits a profile with $|\Omega_1|=0$---every pairwise check passes---whose sources still cannot be merged, a conflict an edge-by-edge view cannot even express. This is the main reason we work with a sheaf at all.

There is also a practical dividend. The coboundary description carries over to settings where an edge-by-edge check would have to be rebuilt by hand: sheaves of partial or interval orders, whose intransitivity can obstruct gluing in ways a single total order cannot; judgment aggregation, which adds logical constraints; and weighted or higher-dimensional complexes, which change the underlying space. The same construction applies in each case without redesign, which is what we mean when we say the sheaf makes the edge check composable.

\subsection{Stochastic Model: Mallows Distribution}

To study how $|\Omega_1|$ varies with preference distributions, we employ the Mallows model \cite{Mallows1957}, a probability distribution over permutations.

\begin{definition}[Mallows Model]
Given a reference permutation $\pi_0$ and dispersion parameter $\phi \in (0,1]$, the Mallows model assigns probability:
$$ P(\pi \mid \pi_0, \phi) \propto \phi^{d(\pi, \pi_0)} $$
where $d(\pi, \pi_0)$ is the Kendall tau distance (number of pairwise inversions) between $\pi$ and $\pi_0$.
\end{definition}

When $\phi \to 0$, the distribution concentrates on $\pi_0$. When $\phi = 1$, the distribution is uniform over all permutations. This allows us to interpolate smoothly between consensus ($\phi$ small) and maximum diversity ($\phi = 1$).

\section{Computational Experiments}

In this section, we analyze several examples to demonstrate the capabilities of our framework. All results were obtained using Python implementations within the Discrete Order Sheaf framework. Basic examples validate the framework's ability to detect known paradoxes (Table~\ref{tab:obstruction_results}).

\begin{table}[ht]
\centering
\caption{Obstruction Loci of Discrete Order Sheaves: Computational Results for Basic Examples. The $|\Omega_2|$ column is identically zero because these interaction graphs have no cells of dimension $\geq 2$ (Section~\ref{sec:obstruction}); it is shown only to make the dimensional vanishing explicit. Detailed configurations for Partial Visibility and $K_4$ are provided in Appendix~\ref{appendix:configurations}.}
\begin{tabular}{lcccp{5cm}}
\toprule
\textbf{Example} & $H^0$ exists & $|\Omega_1|$ & $|\Omega_2|$ & \textbf{Notes} \\
\midrule
Condorcet Triangle & False & 3 & 0 & All 3 edges incompatible \\
Transitive Triangle & True & 0 & 0 & Unanimous agreement \\
Partial Visibility & False & 1 & 0 & Single conflicting edge \\
Complete $K_4$ & False & 5 & 0 & Edge $(V_1, V_4)$ compatible$^*$ \\
\bottomrule
\end{tabular}
\label{tab:obstruction_results}

\vspace{0.3cm}
\footnotesize{$^*$In the $K_4$ example, voters $V_1$ and $V_4$ both have order $A > B > C$, making their shared edge compatible. The remaining 5 edges are incompatible: the three Condorcet edges among $V_1,V_2,V_3$, plus edges $V_2V_4$ and $V_3V_4$, on which $V_4$ (a copy of $V_1$) inherits $V_1$'s two conflicts.}
\end{table}

\subsection{Localization and the Relation to HodgeRank}
\label{sec:results_localization}

We should be precise about how the Obstruction Locus relates to HodgeRank \cite{Jiang2011}, since the two are easily conflated. They operate on \emph{different graphs} and answer \emph{different questions}.

HodgeRank works on the \emph{comparison graph}, whose vertices are the alternatives and whose edges are the compared pairs; the data form a flow $Y$ on these edges, one cardinal margin per pair. Its inconsistency object is the residual $r = Y - \mathrm{grad}\,s$, the part of the flow that no global ranking explains. This residual is itself a function \emph{on edges}: it assigns a magnitude to each pair of alternatives, so HodgeRank already localizes inconsistency, to pairs of \emph{alternatives}. The single integer $\dim H^1$ is something weaker, the dimension of the space in which globally cyclic (harmonic) inconsistency can live. It is a property of the comparison graph, not a reading of the data, and should not be taken for HodgeRank's output.

The Obstruction Locus lives on the \emph{interaction graph}, whose vertices are the sources and whose edges are the pairs required to agree. $\Omega_1(\sigma)$ marks the edges, that is the pairs of \emph{sources}, whose local orders disagree on their shared alternatives. It localizes conflict in the \emph{interaction} structure and identifies \emph{which agents fail to cohere}, a question HodgeRank does not pose: once comparisons are aggregated into one flow, the identity of the individual sources is gone.

Two further differences are intrinsic. First, $\Omega_1$ is purely ordinal: it depends only on the relative orders and is invariant under any monotone re-cardinalization of the data, whereas $r$ and its norm depend both on how ordinal preferences are turned into numerical margins and on the choice of $2$-cells used to separate curl from harmonic. Second, $\Omega_1$ is exact and combinatorial, an integer certificate per edge, while HodgeRank returns continuous residuals that need a threshold before they read as ``conflicting.''

\begin{figure}[ht]
    \centering
    \includegraphics[width=0.9\textwidth]{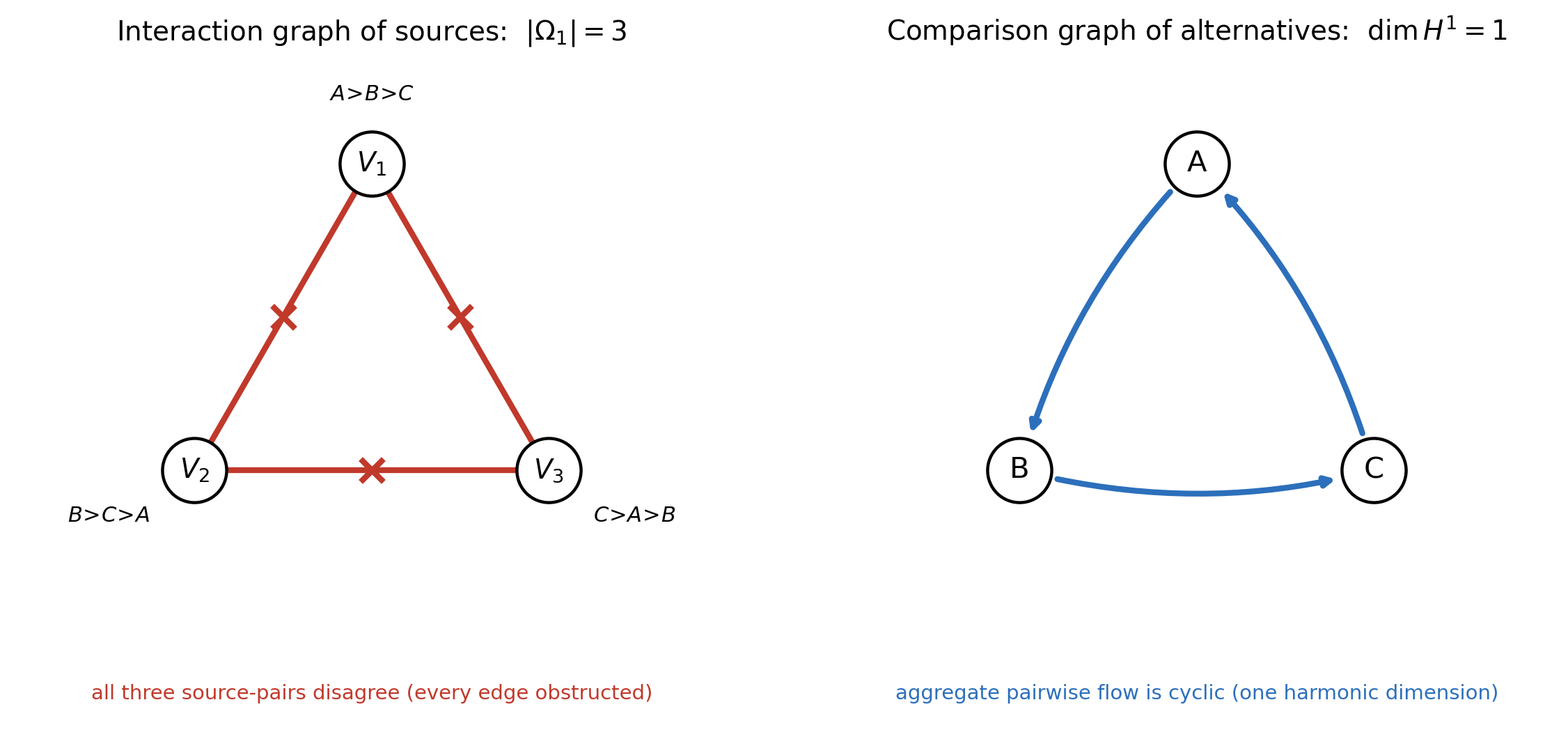}
    \caption{The Obstruction Locus and HodgeRank on the Condorcet triangle, each on its own graph. \emph{Left:} the interaction graph of three sources; every source-pair disagrees on its shared alternatives, so all edges are obstructed ($|\Omega_1|=3$). \emph{Right:} the comparison graph of the three alternatives; the aggregated pairwise flow is cyclic and occupies the single harmonic dimension of the unfilled triangle ($\dim H^1=1$). The two integers count different things on different graphs and are not directly comparable. HodgeRank's data-level localization is carried by the residual edge flow, not by $\dim H^1$; the discrete method's contribution is to localize, ordinally and exactly, in the \emph{interaction} structure.}
    \label{fig:discrete_vs_linearized}
\end{figure}

Figure~\ref{fig:discrete_vs_linearized} illustrates the distinction on the Condorcet triangle, where three sources hold $A>B>C$, $B>C>A$, and $C>A>B$ on the common set $\{A,B,C\}$. On the interaction graph all three source-pairs disagree, so $|\Omega_1| = 3$. On the comparison graph the aggregate is a pure cyclic flow on the triangle of alternatives; with the triangle left unfilled this flow is harmonic and $\dim H^1 = 1$, whereas filling the triangle reclassifies it as curl with $\dim H^1 = 0$. The point is not that $3 > 1$, since these count different things on different graphs, but that the methods answer different questions: HodgeRank asks which \emph{comparisons} are cyclically inconsistent, $\Omega_1$ asks which \emph{interactions} fail to cohere.

\subsection{Consistency of Random Preferences}
\label{sec:random_preferences}

We simulated random preference profiles (over three alternatives) on various graph structures ($N=200{,}000$ trials each; see Appendix~\ref{appendix:configurations} for topology definitions).  We measured two statistical metrics:

\begin{enumerate}
    \item \textbf{Consistency Rate:} The proportion of trials where a global section exists ($H^0 \neq \emptyset$). This measures the probability of spontaneous consensus.
    \item \textbf{Average Incompatibility Index:} The mean value of $|\Omega_1|$ across all trials. This quantifies the expected number of conflicting interactions.
\end{enumerate}

\textbf{Theoretical baseline:} For $|\mathcal{A}|=3$ alternatives, the probability that two random voters agree on their shared ranking is $1/3! = 1/6$. Consequently, an edge is obstructed with probability $5/6$. On a triangle graph ($K_3$), the condition for global consistency requires agreement on two edges (which implies the third by transitivity), yielding a theoretical consistency rate of $(1/6)^2 \approx 2.8\%$.

\begin{figure}[ht]
    \centering
    \includegraphics[width=0.9\textwidth]{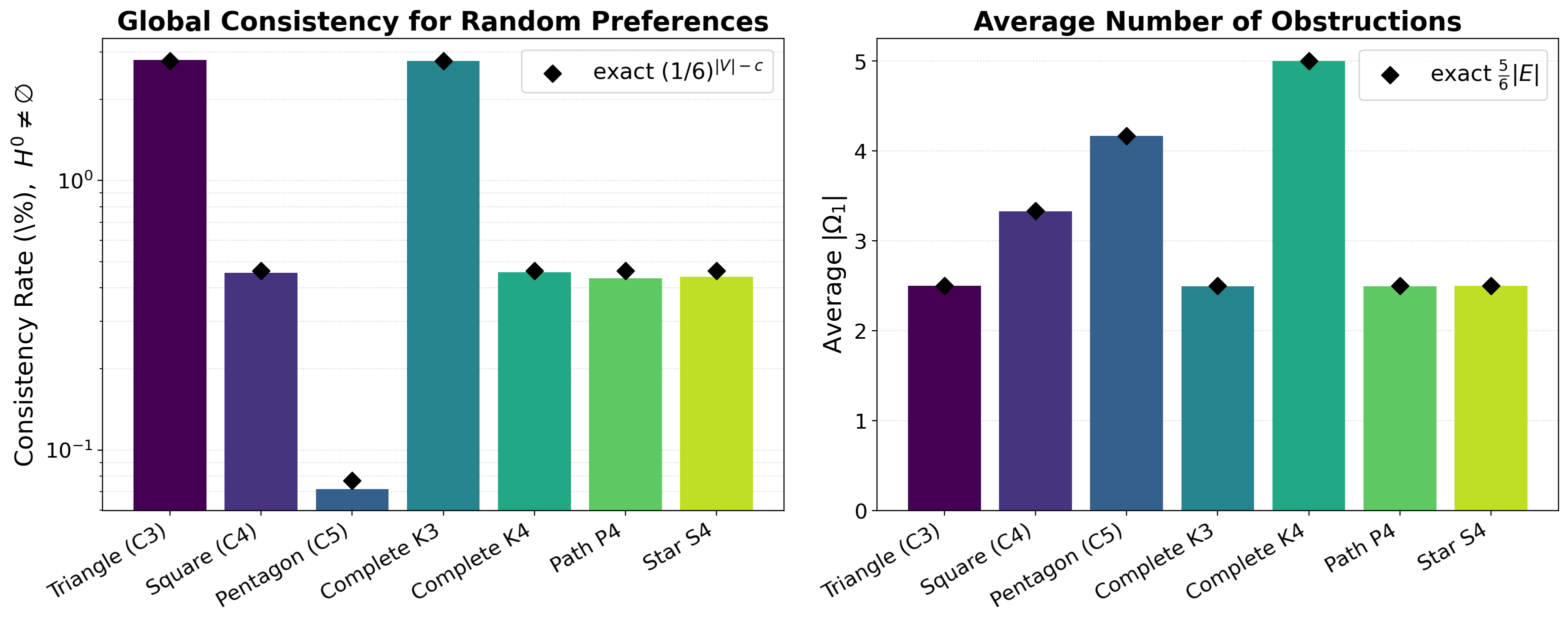}
    \caption{Consistency rate (left, log scale) and mean Incompatibility Index (right) for uniform-random full-visibility profiles on several topologies ($N=200{,}000$ trials each, $|\mathcal{A}|=3$; black diamonds mark the exact values, not fits). \emph{Right:} the mean $|\Omega_1|$ matches the per-edge law $\tfrac{5}{6}|E|$, so the graphs with the most edges ($K_4$, then $C_5$) have the largest mean. \emph{Left:} the consistency rate follows $(1/6)^{|V|-c}$ (with $c$ connected components) and is therefore set by the number of \emph{voters}, not edges: $C_3$ and $K_3$ (both $|V|=3$) coincide at $\approx2.8\%$, the four $|V|=4$ graphs ($C_4,K_4,P_4,S_4$) coincide at $\approx0.46\%$ regardless of their $3$--$6$ edges, and the sparse $C_5$ ($|V|=5$) is lowest. Edge count drives $|\Omega_1|$; vertex count drives consistency, so the dense $K_4$ is \emph{not} the least consistent topology.}
    \label{fig:random_preferences}
\end{figure}

Across topologies the mean Incompatibility Index matches the per-edge prediction $\tfrac{5}{6}|E|$ closely (Figure~\ref{fig:random_preferences}, right), the robust check of the implementation: the graphs with the most edges, $K_4$ ($|E|=6$) and $C_5$ ($|E|=5$), carry the highest mean $|\Omega_1|$, exactly as $\tfrac{5}{6}|E|$ predicts. The consistency rate stands on a different footing, and the two metrics should not be conflated. Under full visibility a profile is globally consistent only when every voter in a connected component holds the \emph{identical} order, so the rate is $(1/6)^{|V|-1}$ (Section~\ref{sec:domain_restrictions}): an all-or-nothing event on the whole component, governed by the number of \emph{voters} and not by the number of edges. The mean $|\Omega_1|$ therefore grows with $|E|$ while the consistency rate does not: the four $|V|=4$ graphs ($C_4, K_4, P_4, S_4$) share the same expected rate $(1/6)^3\approx0.46\%$ despite ranging from three to six edges, and $C_5$ ($|V|=5$) is lower still at $(1/6)^4\approx0.08\%$, even though it is sparser than $K_4$. With $N=200{,}000$ trials the Monte-Carlo estimates sit essentially on these exact values (markers in Figure~\ref{fig:random_preferences}): $C_3$ and $K_3$, being the same graph, coincide at $\approx2.8\%$, the four $|V|=4$ graphs coincide at $\approx0.46\%$, and $C_5$ is the lowest. The left panel is thus a direct picture of the $(1/6)^{|V|-1}$ law, not a ranking of topologies by connectivity; the dense $K_4$ is emphatically \emph{not} the least consistent.

\subsection{Variation of the Obstruction Locus}

We investigate how the Obstruction Locus changes as preference profiles vary. We present two complementary experiments: (1) a stochastic interpolation using Mallows distributions, in which averaging over random draws turns discrete preference switches into a smooth response, and (2) an illustrative deterministic family that serves as scaffolding to show how $|\Omega_1|$ responds to discrete preference changes.

\subsubsection{Stochastic Interpolation via Mallows Model}
\label{sec:stochastic_interpolation}

To demonstrate how $|\Omega_1|$ varies with preference distributions, we constructed a stochastic interpolation model. We consider three voters on a triangle graph, where each voter's preferences are drawn from Mallows distributions.

\textbf{Model setup:} At parameter $t=0$, each voter has a distinct reference ordering that together form a Condorcet cycle: $V_1$ prefers $A>B>C$, $V_2$ prefers $B>C>A$, and $V_3$ prefers $C>A>B$. As $t$ increases toward 1, voters' reference orderings shift toward a common consensus ($A>B>C$) at two discrete switch points (specified below), and the dispersion parameter $\phi$ decreases (concentrating preferences near their references).

\textbf{Interpolation scheme:} The reference ordering for each voter $V_i$ transitions deterministically: $V_2$'s reference changes from $B>C>A$ to $A>B>C$ at $t=0.5$, and $V_3$'s reference changes from $C>A>B$ to $A>B>C$ at $t=0.75$. The dispersion parameter follows $\phi(t) = 0.8 - 0.7t$, so that $\phi(0) = 0.8$ (high diversity) and $\phi(1) = 0.1$ (concentrated near consensus). The smoothness of the $|\Omega_1|$ curve is a consequence of averaging over the stochastic Mallows draws: the underlying reference orderings switch discretely, but the expected number of conflicting edges varies continuously.

We sample $N=500$ profiles at each parameter value and compute the distribution of $|\Omega_1|$.

\begin{figure}[ht]
    \centering
    \includegraphics[width=0.95\textwidth]{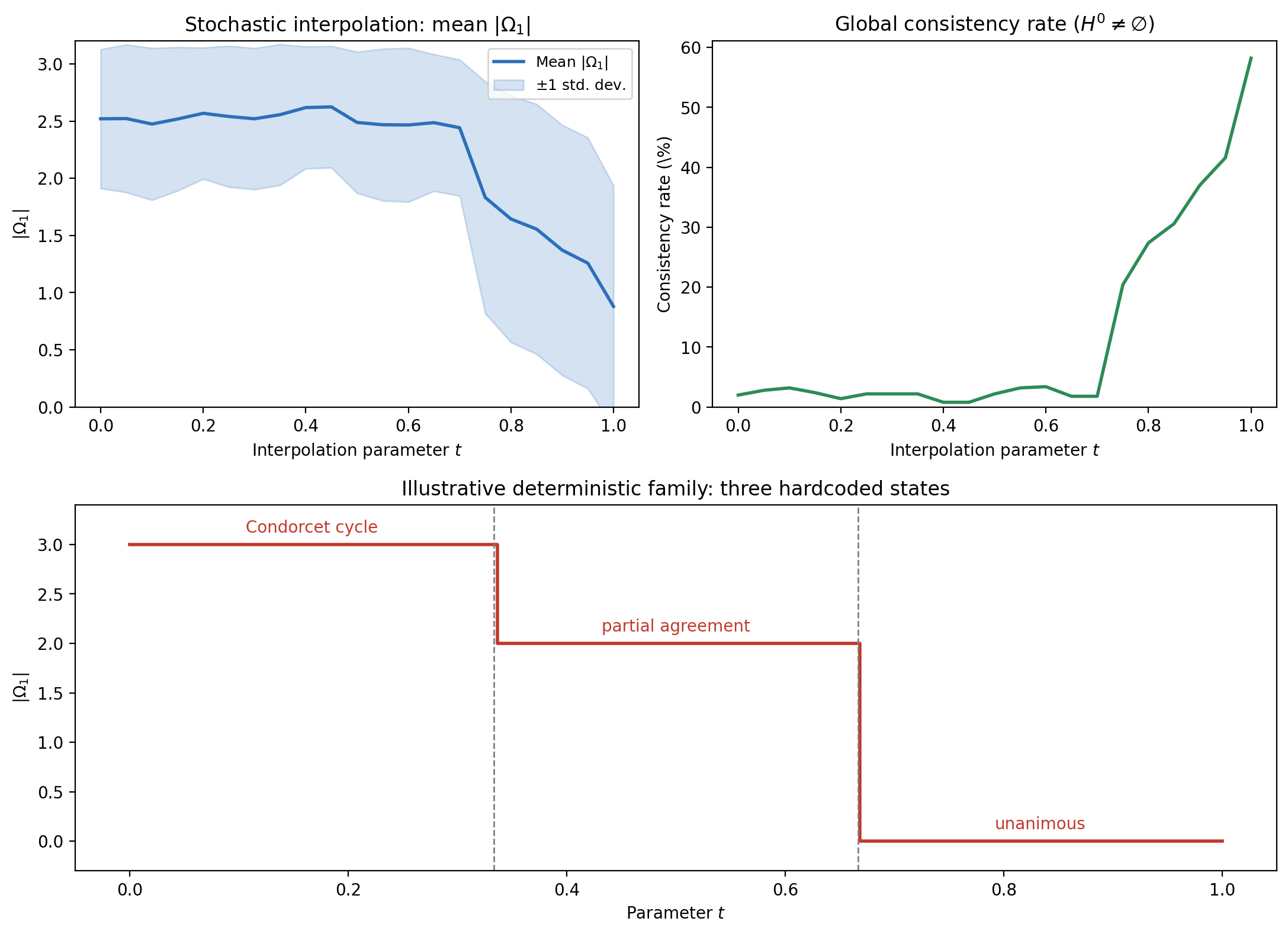}
    \caption{Stochastic interpolation via the Mallows model ($N=500$ draws per $t$). \emph{(Top left)} Mean $|\Omega_1|$ with $\pm1$ standard-deviation band; averaging over the draws yields a smooth decrease from $\approx 2.5$ to $\approx 0.9$. \emph{(Top right)} The consistency rate (probability that $H^0$ exists) rises from a few percent to about $58\%$. \emph{(Bottom)} The illustrative deterministic family of Section~\ref{sec:illustrative_family}, whose $|\Omega_1|$ steps $3\to2\to0$ at the two hardcoded transitions ($t=1/3,2/3$); these steps are programmed by construction, not emergent, and are shown only for contrast with the smooth stochastic curve above.}
    \label{fig:stochastic_interpolation}
\end{figure}

\textbf{Results} (Figure~\ref{fig:stochastic_interpolation}):
\begin{enumerate}
    \item The mean $|\Omega_1|$ transitions smoothly from $2.52 \pm 0.61$ at $t=0$ to $0.88 \pm 1.06$ at $t=1$ (seed-dependent at the level of the last digit; $N=500$).
    \item The standard deviation grows as preferences concentrate, peaking near $t=1$, where individual profiles range from fully cyclic to fully consistent and $|\Omega_1|$ is most variable.
    \item The consistency rate (probability that $H^0$ exists) rises overall (modulo sampling noise, which leaves it non-monotonic in the high-dispersion regime) from about $2\%$ to about $58\%$.
    \item The distribution of $|\Omega_1|$ values shifts continuously from being concentrated at high values (2-3) to low values (0-1).
\end{enumerate}

These results show that $|\Omega_1|$ responds continuously and predictably to gradual changes in the preference distribution.

\subsubsection{Illustrative Deterministic Family}
\label{sec:illustrative_family}

To illustrate the framework's sensitivity to preference changes, we also constructed a deterministic parametric family $\{F_t\}_{t \in [0,1]}$ with hardcoded transitions at $t=1/3$ and $t=2/3$, where voters switch their preferences according to a predefined schedule.

\textbf{Important caveat:} This experiment is illustrative scaffolding, not evidence of a genuine transition. The discrete jumps in $|\Omega_1|$ at $t=1/3$ and $t=2/3$ occur by design: we explicitly programmed voters to change their preferences at these points. The experiment serves only to verify that our computational framework correctly detects these changes and to provide a visual contrast with the smooth stochastic response.

The results (Figure~\ref{fig:stochastic_interpolation}) show $|\Omega_1|$ dropping from 3 to 2 at $t=1/3$, and from 2 to 0 at $t=2/3$, exactly as constructed.

\subsection{The Pushforward under Quotient}

\begin{figure}[ht]
    \centering
    \includegraphics[width=0.9\textwidth]{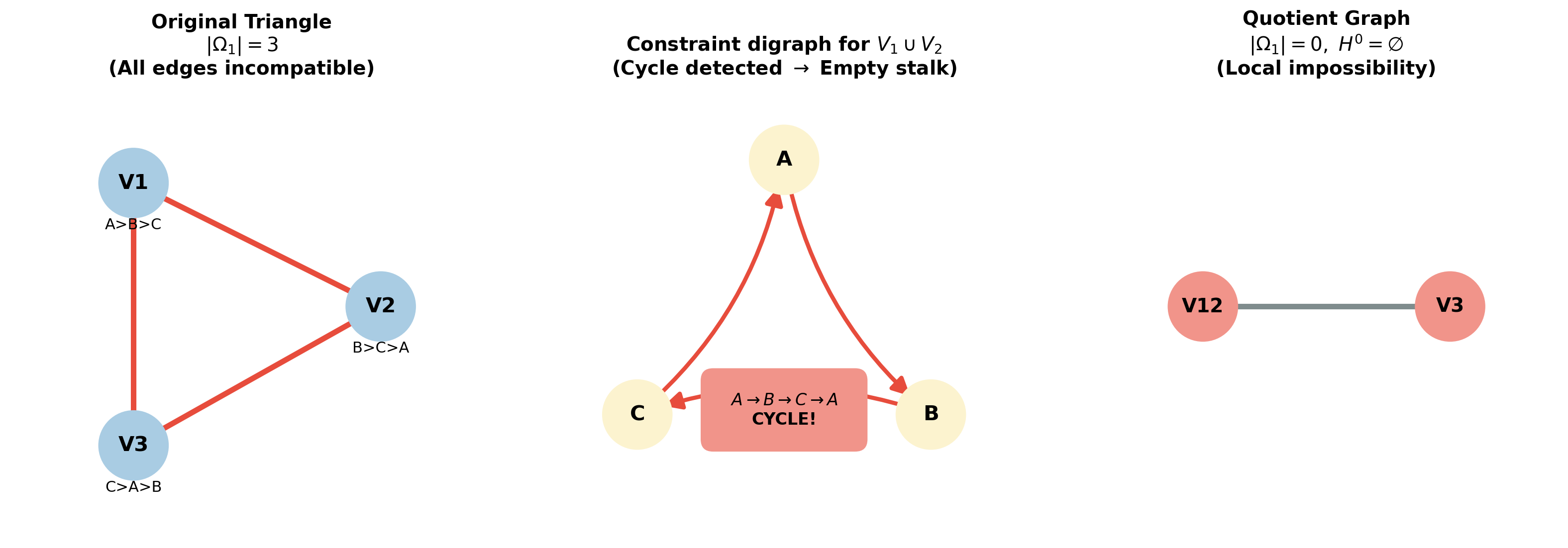}
    \caption{Pushforward under a vertex merge. \emph{(Left)} The Condorcet triangle, $|\Omega_1|=3$. \emph{(Middle)} Merging $V_1$ and $V_2$ builds the constraint digraph on $\{A,B,C\}$; the cycle $A\to B\to C\to A$ certifies that the merged stalk is empty (Proposition~\ref{prop:empty_iff_cycle}). \emph{(Right)} On the quotient graph the edge Incompatibility Index drops to $|\Omega_1|=0$, yet no global section exists ($H^0=\emptyset$): the obstruction has moved from the edges to the empty stalk at $V_{12}$.}
    \label{fig:pushforward_rigorous}
\end{figure}

Let us examine the behavior of the sheaf under a graph quotient map $\pi$. We start with a Condorcet cycle (V1, V2, V3), where $|\Omega_1|=3$, and merge V1 and V2 into V12 (Figure~\ref{fig:pushforward_rigorous}). Then we apply the pushforward $\pi_* F$ using the constraint digraph method described in Section~\ref{sec:pushforward}.

To compute the stalk $(\pi_* F^\sigma)_{V12}$, we build a constraint digraph from the orders of V1 ($A>B>C$) and V2 ($B>C>A$):
\begin{itemize}
    \item V1 imposes: $A \to B$, $B \to C$
    \item V2 imposes: $B \to C$, $C \to A$
\end{itemize}
The resulting digraph contains edges $A \to B \to C \to A$, forming a directed cycle. This cycle immediately certifies that no total order can satisfy both constraints, proving that the stalk $(\pi_* F^\sigma)_{V12}$ is \textit{empty}. See Appendix \ref{appendix:proofs} for the formal proof and Appendix \ref{appendix:example} for the complete digraph construction.

The Incompatibility Index on the quotient graph drops to $|\Omega_1|=0$: the single remaining edge $V_{12}V_3$ has an empty endpoint stalk at $V_{12}$, which carries no order to compare, so by the convention of Remark~\ref{rem:omega_convention} it contributes no edge obstruction. However, $H^0$ remains empty due to the local impossibility at V12. The obstruction transformed from edge incompatibilities into the structure of the stalk itself. This empty-stalk phenomenon is verified computationally via the constraint digraph cycle, not merely proven abstractly. Appendix~\ref{appendix:example2} gives the complementary case, in which the base profile has \emph{no} obstructed edges at all ($|\Omega_1|=0$) yet the merge is still impossible---an obstruction no edge-by-edge check can see.

\subsection{Scaling Experiments}
\label{sec:scaling}

To validate the practical efficiency of the digraph-based pushforward algorithm, we conducted scaling experiments measuring runtime and memory usage across different problem sizes.

\subsubsection{Scaling with Number of Alternatives}

We tested the cycle detection algorithm for $|\mathcal{A}| \in \{6, 8, 10, 12\}$ alternatives with $|P|=5$ voters in the merge preimage (Table~\ref{tab:scaling}). For $|\mathcal{A}| \leq 8$, we measured naive enumeration time directly; for larger $|\mathcal{A}|$, naive times are extrapolated since direct measurement is infeasible.

\begin{table}[ht]
\centering
\caption{Complexity Comparison: Constraint Digraph vs Naive Enumeration}
\begin{tabular}{rrrrrr}
\toprule
$|\mathcal{A}|$ & $|\mathcal{A}|!$ & Digraph (ms) & Naive (ms) & Speedup & Memory (KB) \\
\midrule
6 & 720 & 0.53 & 0.20 & $<1\times$ & 18.9 \\
8 & 40,320 & 0.75 & 13.6 & 18$\times$ & 20.9 \\
10 & 3,628,800 & 0.97 & 1,222$^*$ & 1,263$\times$ & 25.3 \\
12 & 479,001,600 & 1.08 & 161,316$^*$ & 149,300$\times$ & 30.1 \\
\bottomrule
\end{tabular}
\label{tab:scaling}

\vspace{0.2cm}
\footnotesize{$^*$Estimated by extrapolation from $|\mathcal{A}|=8$. Timing averaged over 50 trials, $|P|=5$ voters.}
\end{table}

For small $|\mathcal{A}|=6$, the overhead of building the digraph exceeds the naive enumeration time. However, the crossover occurs around $|\mathcal{A}|=7$, after which the digraph approach dominates. At $|\mathcal{A}|=12$, the speedup exceeds $149{,}000\times$: naive enumeration would require approximately 2.7 minutes versus $\sim$1 ms for digraph cycle detection.

Memory usage scales modestly: from 18.9 KB at $|\mathcal{A}|=6$ to 30.1 KB at $|\mathcal{A}|=12$, reflecting the $O(|\mathcal{A}|^2)$ edge storage in the constraint digraph.

\subsubsection{Realistic Scenario: Committee Decision}

We simulated a realistic preference aggregation scenario with $|V|=50$ voters and $|\mathcal{A}|=8$ alternatives on a random graph (Erd\H{o}s-R\'enyi, $p=0.15$, yielding 179 edges). Computing $\Omega_1$ over all edges required 0.15 ms. The incompatibility rate was 97.2\% (174 of 179 edges), reflecting a high-conflict scenario with diverse preferences.

For pushforward computation, we merged 5 randomly selected voters. Cycle detection completed in 1.9 ms with 26.0 KB memory, confirming an empty stalk via the cycle $(0 \to 2 \to 1 \to 0)$. Both $\Omega_1$ and the pushforward thus run in at most a few milliseconds at this scale.

\subsubsection{Scaling with Merge Size}

We also tested how cycle detection time scales with the number of voters being merged ($|P|$), holding $|\mathcal{A}|=8$ fixed. Runtime increased from 0.34 ms at $|P|=3$ to 15.5 ms at $|P|=50$, consistent with the $O(|\mathcal{A}|^2 \cdot |P|)$ complexity bound.

The conflict rate was 100\% across all merge sizes when using uniformly random preferences: random voters almost always have conflicting constraints, resulting in empty stalks.

\section{Discussion}

The computational framework presented here establishes a foundation for analyzing social choice obstructions using discrete sheaf theory. By focusing on localization of incompatibilities, we have developed metrics and tools that accurately reflect the structure of aggregation paradoxes.

\subsection{Interpretation of Results}

The distinction between the Incompatibility Index $|\Omega_1|$ and the linearized cohomology $H^1_{\rm lin}$---the first cohomology of HodgeRank's comparison complex, whose dimension counts the harmonic (globally cyclic) inconsistencies (Section~\ref{sec:results_localization})---is significant (cf. Remark~\ref{remark:what_omega_measures}). Whereas $\dim H^1_{\rm lin}$ is a property of the comparison graph rather than a reading of the data, $|\Omega_1|$ localizes disagreement in the interaction structure, ordinally and with an exact per-edge certificate \cite{Jiang2011}. This localization is what shows how specific interactions contribute to global impossibilities.

The stochastic interpolation experiment (Section~\ref{sec:stochastic_interpolation}) shows that $|\Omega_1|$ responds smoothly to changes in the preference distribution: averaging over Mallows draws turns the discrete reference switches into a continuous response of the mean, with the spread of $|\Omega_1|$ largest once preferences concentrate near $t=1$.

The most significant experimental finding is the behavior of the pushforward ($\pi_*$). The empty stalk phenomenon reveals a fundamental principle: aggregation conflicts cannot be resolved by simple merging or coarse-graining without information loss or the imposition of an aggregation rule. The pushforward functor captures this impossibility by transforming distributed conflicts (edge obstructions) into localized impossibilities (empty stalks), tracking the obstruction across different scales of analysis.

\subsection{Connection to Classical Social Choice Theory}

This framework offers a new lens through which to view classical results.

\subsubsection{Domain Restrictions}
\label{sec:domain_restrictions}
Classical social choice theory identifies specific restrictions on the domain of preferences that guarantee consistent aggregation, such as single-peaked preferences \cite{Black1948}. Single-peakedness typically implies an underlying linear structure, and in deliberative settings such structure can emerge as meta-agreement among participants on a common dimension of evaluation \cite{List2002}. We caution against a tempting but incorrect reading of our random-profile experiment in this connection. For fully visible uniform profiles a global section exists only when all voters in a connected component hold the identical order, so the consistency rate is $(1/6)^{|V|-1}$: it is governed by the number of voters, \emph{not} by whether the interaction graph is acyclic. What acyclicity does control is the \emph{number} of obstructions: with fewer edges, acyclic graphs have a lower mean $|\Omega_1|$, but not a higher consistency rate. A genuine domain restriction such as single-peakedness instead constrains \emph{which} orders occur, and the sheaf framework lets one study how such constraints force the Obstruction Locus to vanish ($\Omega=0$); we leave a quantitative treatment to future work.

\subsubsection{Connection to Judgment Aggregation}
The framework naturally extends to judgment aggregation settings \cite{ListPettit2002}. In judgment aggregation, individuals hold positions on logically connected propositions, and the challenge is to aggregate these into a consistent collective judgment. The Obstruction Locus can identify which logical constraints are violated when aggregating, providing localized diagnostics analogous to our preference setting.

\subsection{Computational Complexity}

The computational tractability of this framework is crucial for practical applications.

\subsubsection{Computing the Obstruction Locus}

Computing the Obstruction Locus $\Omega_1$ involves checking compatibility on every edge. If $|\mathcal{A}|$ is the number of alternatives, determining the restriction on an overlap takes $O(|\mathcal{A}|)$ time. Thus, the total complexity for computing $\Omega_1$ is $O(|E| \cdot |\mathcal{A}|)$, which is polynomial and efficient (Appendix \ref{appendix:algorithms}).

\subsubsection{Computing the Pushforward}

The pushforward reduces to building a constraint digraph and detecting a cycle (Algorithm~\ref{alg:pushforward_dag}). Cycle detection---hence empty-stalk verification---runs in $O(|\mathcal{A}| \cdot |P| + |E_C|)$ time, where $|E_C| \leq |\mathcal{A}| \cdot |P|$ is the number of constraint edges, simplifying to $O(|\mathcal{A}| \cdot |P|)$. The $O(|\mathcal{A}|^2 \cdot |P|)$ bound stated in Section~\ref{sec:pushforward} is conservative, covering the case where constraint edges approach $|\mathcal{A}|^2$ (many voters with diverse preferences).

\begin{table}[ht]
\centering
\caption{Pushforward Complexity Comparison}
\begin{tabular}{lcc}
\toprule
\textbf{Approach} & \textbf{Complexity} & \textbf{Feasible for $|\mathcal{A}|$} \\
\midrule
Naive (factorial) & $O(|\mathcal{A}|! \cdot |P|)$ & $\leq 8$ \\
Digraph (cycle detection) & $O(|\mathcal{A}|^2 \cdot |P|)$ & $\leq 100+$ \\
Digraph (count all orders) & \#P-complete & varies \\
\bottomrule
\end{tabular}
\end{table}

\textbf{Caveat on counting vs.\ enumeration:} Cycle detection---hence empty-stalk detection and existence proofs---is polynomial, but \emph{counting} all compatible orders is \#P-complete (Remark~\ref{rem:counting}). Our implementation therefore returns a single witness ordering for large $|\mathcal{A}|$.

For $|\mathcal{A}| = 10$ alternatives and $|P| = 3$ voters:
\begin{itemize}
    \item Naive: $10! \cdot 3 \approx 10.9$ million operations
    \item Digraph (cycle detection): $\sim 300$ operations
\end{itemize}

This dramatic speedup makes the pushforward computation practical for real-world applications where detecting impossibility (empty stalks) is the primary goal.

\subsection{Limitations and Future Directions}
\label{sec:limitations}

Several limitations warrant discussion:

\textbf{Triviality of higher obstructions.} On a graph the obstruction loci $\Omega_k$ ($k\geq 2$) vanish for dimensional reasons. They would also stay empty under our \v{C}ech construction on a higher-dimensional complex, because a triple overlap is contained in each pairwise overlap, so agreement on edges forces agreement on the triple---this is a feature of the construction rather than something special to total orders, and it holds for partial and interval orders too. Note that this is a statement about agreement on overlaps only: it does \emph{not} say that pairwise-consistent local orders glue into a global ranking, and indeed they need not (the empty-stalk phenomenon of the pushforward). Capturing meaningful higher-dimensional obstructions therefore requires a finer construction together with higher cells, a direction we leave to future work.

\textbf{Graph selection.} The framework assumes a known, fixed interaction graph $G$. In practice, the choice of $G$, i.e.\ which voter pairs are required to be consistent, is a modeling decision that significantly affects results. Future work should address principled methods for constructing $G$ from data.

\textbf{Scalability.} While our algorithms are polynomial, computing $\Omega_1$ for very large graphs ($|V| \gg 1000$) with many alternatives may require parallelization or approximation schemes.

\textbf{Future directions.} Connecting this framework to \textbf{persistent homology} \cite{Bubenik2014} could enable analysis of how preference profiles evolve over time or across different scales of interaction. Extending to weighted graphs (where edges have varying importance) would also increase practical applicability.

\section{Conclusion}

We have introduced a computational framework for analyzing discrete preference aggregation using sheaf theory. By defining the Obstruction Locus ($\Omega$), we provide a tool to localize and quantify the impediments to global consistency. Our experiments demonstrate that $|\Omega_1|$ varies smoothly with preference distributions (via stochastic interpolation using Mallows models) and that the pushforward transforms edge conflicts into local impossibilities (empty stalks). The polynomial-time pushforward algorithm, based on constraint digraph cycle detection, makes these methods practical for real-world applications.

The sheaf-theoretic language provides a clean framework for this localization problem, enabling precise definitions and the pushforward construction. We hope this framework proves useful for analyzing preference aggregation in settings where knowing where conflicts occur matters as much as knowing that they occur.

\section*{Competing Interests}
The author declares no competing interests.
 
% Appendices
\appendix

\section{Algorithmic Details}
\label{appendix:algorithms}

\subsection{Computing the Obstruction Locus ($\Omega_1$)}

The Obstruction Locus $\Omega_1(\sigma)$ is computed by iterating over all edges and checking the compatibility of the restricted local orders.

\begin{algorithm}[H]
\caption{Compute $\Omega_1(\sigma)$}
\label{alg:omega1}
\begin{algorithmic}[1]
\Require Graph $G=(V,E)$, Preference profile $\sigma$, Visibility sets $\{\mathcal{A}_v\}_{v\in V}$
\Ensure Obstruction locus $\Omega_1$
\State $\Omega_1 \gets \emptyset$
\For{each edge $e=\{u,v\} \in E$}
    \State $\mathcal{A}_{uv} \gets \mathcal{A}_u \cap \mathcal{A}_v$ \Comment{Compute overlap}
    \If{$|\mathcal{A}_{uv}| \geq 2$}
        \State $\sigma_u^e \gets$ restriction of $\sigma_u$ to $\mathcal{A}_{uv}$
        \State $\sigma_v^e \gets$ restriction of $\sigma_v$ to $\mathcal{A}_{uv}$
        \If{$\sigma_u^e \neq \sigma_v^e$}
            \State $\Omega_1 \gets \Omega_1 \cup \{e\}$ \Comment{Incompatible edge found}
        \EndIf
    \EndIf
\EndFor
\Return $\Omega_1$
\end{algorithmic}
\end{algorithm}

\noindent\textbf{Complexity:} $O(|E| \cdot |\mathcal{A}|)$.

\subsection{Pushforward Computation via Constraint Digraph}

Given a quotient map $\pi: G \to G'$, we compute the stalk of the pushforward sheaf $(\pi_* F^\sigma)_{v'}$ at a vertex $v' \in G'$ using a polynomial-time constraint digraph approach.

\begin{algorithm}[H]
\caption{Compute Stalk of Pushforward $(\pi_* F^\sigma)_{v'}$ (Constraint Digraph Method)}
\label{alg:pushforward_dag}
\begin{algorithmic}[1]
\Require Quotient map $\pi$, Vertex $v' \in G'$, Profile $\sigma$ on $G$, Alternatives $\mathcal{A}$
\Ensure Stalk $S_{v'}$ (Set of compatible orders, or $\emptyset$ if empty)
\State $P \gets \pi^{-1}(v')$ \Comment{Identify preimage in G}
\State Initialize directed graph $C$ with vertices $\mathcal{A}$ and no edges \Comment{Constraint digraph}
\State
\For{each $v \in P$} \Comment{Build constraint digraph}
    \State $\sigma_v \gets$ local order at vertex $v$
    \State $\mathcal{A}_v \gets$ visibility set at $v$
    \For{each consecutive pair $(a_i, a_{i+1})$ in $\sigma_v$}
        \State Add directed edge $a_i \to a_{i+1}$ to $C$ \Comment{Pairwise constraint}
    \EndFor
\EndFor
\State
\If{$C$ contains a directed cycle} \Comment{Detect contradictory constraints}
    \State \Return $\emptyset$ \Comment{Empty stalk (local impossibility)}
\Else
    \State $S_{v'} \gets$ \{all topological orderings of $C$\} \Comment{All compatible orders}
    \State \Return $S_{v'}$
\EndIf
\end{algorithmic}
\end{algorithm}

\noindent\textbf{Complexity:} 
\begin{itemize}
    \item \textbf{Digraph construction:} $O(|\mathcal{A}| \cdot |P|)$: each voter contributes $O(|\mathcal{A}|)$ edges
    \item \textbf{Cycle detection:} $O(|\mathcal{A}| + |E_C|)$: depth-first search, where $|E_C| \leq |\mathcal{A}| \cdot |P|$
    \item \textbf{Total:} $O(|\mathcal{A}|^2 \cdot |P|)$: polynomial in both $|\mathcal{A}|$ and $|P|$
\end{itemize}

This is a dramatic improvement over the naive $O(|\mathcal{A}|! \cdot |P|)$ factorial approach.

\begin{remark}[Enumeration of Compatible Orders]
If the constraint digraph is acyclic, the number of compatible orders equals the number of linear extensions of $C$; counting them is \#P-complete (Remark~\ref{rem:counting}), but cycle detection alone suffices for empty-stalk verification.
\end{remark}

\section{Worked Example: Pushforward Computation via Constraint Digraph}
\label{appendix:example}

Consider the Condorcet triangle on $G$ with alternatives $\mathcal{A}=\{A, B, C\}$:
\begin{align*}
\sigma_{V1} &: A > B > C \\
\sigma_{V2} &: B > C > A \\
\sigma_{V3} &: C > A > B
\end{align*}

We apply the quotient map $\pi: G \to G'$ that merges V1 and V2 into V12. The preimage is $P = \{V1, V2\}$.

\subsection*{Step 1: Build Constraint Digraph}

We create a directed graph $C$ with vertices $\{A, B, C\}$ and add edges based on each voter's constraints:

\textbf{From V1} ($A > B > C$):
\begin{itemize}
    \item Add edge $A \to B$ (A must precede B)
    \item Add edge $B \to C$ (B must precede C)
\end{itemize}

\textbf{From V2} ($B > C > A$):
\begin{itemize}
    \item Add edge $B \to C$ (already present)
    \item Add edge $C \to A$ (C must precede A)
\end{itemize}

\textbf{Resulting digraph $C$:} Edges are $\{A \to B, B \to C, C \to A\}$.

\subsection*{Step 2: Detect Cycles}

The path $A \to B \to C \to A$ forms a directed cycle. This immediately proves:

\begin{center}
\fbox{\textbf{The stalk $(\pi_* F^\sigma)_{V12}$ is EMPTY.}}
\end{center}

\subsection*{Step 3: Interpretation}

The cycle $A \to B \to C \to A$ encodes the logical contradiction:
\begin{itemize}
    \item V1 requires: $A$ before $B$ before $C$
    \item V2 requires: $C$ before $A$
    \item Combined: $A$ before $B$ before $C$ before $A$ (impossible!)
\end{itemize}

No total order on $\{A, B, C\}$ can simultaneously satisfy both V1's and V2's preferences. The constraint digraph makes this contradiction explicit and verifiable in polynomial time.

\subsection*{Comparison with Naive Approach}

For completeness, we verify by enumeration. We check all $3! = 6$ permutations of $\mathcal{A}$:

\begin{enumerate}
    \item $\tau = (A,B,C)$: Compatible with V1 ($\checkmark$), Incompatible with V2 ($\times$).
    \item $\tau = (A,C,B)$: Incompatible with V1 ($\times$).
    \item $\tau = (B,A,C)$: Incompatible with V1 ($\times$).
    \item $\tau = (B,C,A)$: Incompatible with V1 ($\times$), Compatible with V2 ($\checkmark$).
    \item $\tau = (C,A,B)$: Incompatible with V1 ($\times$).
    \item $\tau = (C,B,A)$: Incompatible with V1 ($\times$).
\end{enumerate}

Since no total order $\tau$ is compatible with both V1 and V2 simultaneously, the stalk $(\pi_* F^\sigma)_{V12}$ is empty. This matches the digraph result but required checking 6 orderings. For $|\mathcal{A}|=10$, the naive approach would require checking $10! = 3{,}628{,}800$ orderings, while the digraph approach needs only cycle detection ($O(|\mathcal{A}|^2) = 100$ operations).

\section{Proofs}
\label{appendix:proofs}

\begin{proposition}[Empty stalk $\Leftrightarrow$ constraint cycle]
\label{prop:empty_iff_cycle}
Let $\pi: G \to G'$ be a quotient map, $v' \in G'$ a vertex with preimage $P = \pi^{-1}(v')$, and $\sigma$ a profile on $G$. Let $C(P)$ be the constraint digraph on vertex set $\mathcal{A}_{v'} = \bigcup_{v\in P}\mathcal{A}_v$ with an edge $a \to b$ whenever $a >_{\sigma_v} b$ for some $v \in P$. Then the pushforward stalk $(\pi_*F^\sigma)_{v'}$ is nonempty if and only if $C(P)$ is acyclic; equivalently, it is empty if and only if $C(P)$ contains a directed cycle. This holds for any size of $P$, whether or not the pairwise overlaps are individually consistent.
\end{proposition}

\begin{proof}
By definition, $(\pi_*F^\sigma)_{v'}$ is the set of total orders $\tau$ on $\mathcal{A}_{v'}$ with $\tau|_{\mathcal{A}_v} = \sigma_v$ for every $v \in P$. For a fixed $v$, the condition $\tau|_{\mathcal{A}_v} = \sigma_v$ holds exactly when $a >_\tau b$ for every pair with $a >_{\sigma_v} b$; ranging over $v \in P$, this is precisely the requirement that $\tau$ respect every edge of $C(P)$. Such a $\tau$ is therefore a linear extension of $C(P)$. A finite digraph admits a linear extension (a topological order) if and only if it is acyclic, so the stalk is nonempty iff $C(P)$ is acyclic. If $C(P)$ contains a cycle $a_1 \to a_2 \to \cdots \to a_k \to a_1$, no order can satisfy $a_1 >_\tau \cdots >_\tau a_k >_\tau a_1$, so the stalk is empty.
\end{proof}

\begin{remark}
The pairwise case is recovered when $|P|=2$: if $\sigma_{V_1}|_{\mathcal{A}_{12}} \neq \sigma_{V_2}|_{\mathcal{A}_{12}}$ on the overlap $\mathcal{A}_{12}=\mathcal{A}_1\cap \mathcal{A}_2$, some pair $a,b\in \mathcal{A}_{12}$ is ordered oppositely by the two voters, producing a $2$-cycle $a \to b \to a$ in $C(P)$ and hence an empty stalk. The proposition also covers the strictly more general situation (worked out in Appendix~\ref{appendix:example2}) in which every pairwise overlap is consistent yet a longer directed cycle makes the merge impossible.
\end{remark}

\section{Worked Example: An Obstruction Invisible to Pairwise Checking}
\label{appendix:example2}

The Condorcet merge of Appendix~\ref{appendix:example} produces an empty stalk from two sources that \emph{already} disagree on their shared alternatives: edge $V_1V_2$ is obstructed in the base graph, so a plain pairwise comparison already flags it. The following example exhibits the genuinely more general situation promised in Section~\ref{sec:why_sheaf}---every pairwise overlap is consistent, so $|\Omega_1|=0$ and a global section of the base sheaf exists, yet merging the sources is impossible.

Consider three sources with partial visibility over $\mathcal{A}=\{A,B,C\}$:
\begin{align*}
V_1 &\text{ sees } \{A,B\}, & \sigma_{V_1} &: A>B,\\
V_2 &\text{ sees } \{B,C\}, & \sigma_{V_2} &: B>C,\\
V_3 &\text{ sees } \{A,C\}, & \sigma_{V_3} &: C>A.
\end{align*}
Each pair of sources shares exactly one alternative, so the interaction graph is the triangle $C_3$ with overlaps $\mathcal{A}_{12}=\{B\}$, $\mathcal{A}_{23}=\{C\}$, $\mathcal{A}_{13}=\{A\}$. Every overlap has size one, so the restricted orders coincide trivially on every edge:
\[
|\Omega_1(\sigma)| = 0, \qquad H^0(G,F)\neq\emptyset.
\]
No edge-by-edge comparison detects any conflict, and the local orders genuinely do cohere pairwise: the profile $\sigma$ is itself a global section of the base sheaf.

Now merge all three sources under the quotient $\pi:G\to G''$ with total preimage $P=\{V_1,V_2,V_3\}$. The constraint digraph on $\mathcal{A}_{v''}=\{A,B,C\}$ collects one edge per source,
\[
A\to B \ (\text{from } V_1),\qquad B\to C \ (\text{from } V_2),\qquad C\to A \ (\text{from } V_3),
\]
which forms the directed cycle $A\to B\to C\to A$. By Proposition~\ref{prop:empty_iff_cycle} the stalk $(\pi_*F^\sigma)_{v''}$ is empty: no total order on $\{A,B,C\}$ extends all three local orders, so the merged sources admit no common ranking.

This is exactly the behaviour an edge-by-edge view cannot express. The base profile has \emph{no} obstructed edge ($|\Omega_1|=0$) and is globally consistent as a section, yet fusion into a single ranking fails; the obstruction is a length-three cycle that becomes visible only when the three constraints are placed in one stalk. It is this case---pairwise-consistent but globally unmergeable---that the pushforward, and not a pairwise check, detects, and that motivates the remark following Proposition~\ref{prop:empty_iff_cycle}.

\section{Configuration of Validation Examples and Experiments}
\label{appendix:configurations}

To ensure reproducibility, we provide the exact graph topologies, visibility sets, and preference assignments used in Section 4.

\subsection{Basic Validation Examples}

The results reported in Table~\ref{tab:obstruction_results} correspond to the following specific configurations.

\paragraph{Partial Visibility Example ($|\Omega_1|=1$)}
This example demonstrates that obstructions can arise solely from the restriction to shared alternatives, even if the graph has cycles.
\begin{itemize}
    \item \textbf{Graph:} Cycle graph $C_3$ (Triangle) with vertices $V=\{V_1, V_2, V_3\}$.
    \item \textbf{Alternatives:} $\mathcal{A}_{global} = \{A, B, C, D\}$.
    \item \textbf{Visibility Sets:}
    \begin{itemize}
        \item $V_1$ sees $\{A, B, C\}$
        \item $V_2$ sees $\{B, C, D\}$
        \item $V_3$ sees $\{A, C, D\}$
    \end{itemize}
    \item \textbf{Preferences (Orders):}
    \begin{itemize}
        \item $\sigma_{V1}: A > B > C$
        \item $\sigma_{V2}: B > C > D$
        \item $\sigma_{V3}: C > D > A$
    \end{itemize}
    \item \textbf{Analysis:}
    \begin{itemize}
        \item Edge $(V_1, V_2)$: Overlap $\{B, C\}$. $V_1$ implies $B>C$; $V_2$ implies $B>C$. \textbf{Compatible.}
        \item Edge $(V_2, V_3)$: Overlap $\{C, D\}$. $V_2$ implies $C>D$; $V_3$ implies $C>D$. \textbf{Compatible.}
        \item Edge $(V_3, V_1)$: Overlap $\{A, C\}$. $V_1$ implies $A>C$; $V_3$ implies $C>A$. \textbf{Incompatible.}
    \end{itemize}
    \item \textbf{Result:} Total $|\Omega_1| = 1$.
\end{itemize}

\paragraph{Complete $K_4$ Example ($|\Omega_1|=5$)}
This example represents a "Condorcet + 1 Consensus" configuration on a tetrahedron.
\begin{itemize}
    \item \textbf{Graph:} Complete graph $K_4$ (4 vertices, 6 edges).
    \item \textbf{Alternatives:} $\{A, B, C\}$ visible to all.
    \item \textbf{Preferences:}
    \begin{itemize}
        \item $\sigma_{V1}: A > B > C$
        \item $\sigma_{V2}: B > C > A$
        \item $\sigma_{V3}: C > A > B$
        \item $\sigma_{V4}: A > B > C$ \quad (Identical to $V_1$)
    \end{itemize}
    \item \textbf{Analysis:} Vertices $V_1, V_2, V_3$ form a Condorcet cycle (3 conflicting edges). Vertex $V_4$ agrees perfectly with $V_1$ (edge $V_1-V_4$ is compatible), but inherits $V_1$'s conflicts with $V_2$ and $V_3$ (2 conflicting edges).
    \item \textbf{Result:} Total $|\Omega_1| = 3 + 0 + 2 = 5$.
\end{itemize}

\subsection{Topologies for Random Preference Experiments}

In Section~\ref{sec:random_preferences} (Figure~\ref{fig:random_preferences}), we evaluated consistency rates on the following graph topologies. In all cases, the set of alternatives was $A=\{A, B, C\}$ and preferences were drawn uniformly at random from the $3! = 6$ possible permutations.

\begin{table}[h]
\centering
\begin{tabular}{lccl}
\toprule
\textbf{Topology} & \textbf{Nodes ($|V|$)} & \textbf{Edges ($|E|$)} & \textbf{Description} \\
\midrule
Triangle ($C_3$) & 3 & 3 & Cycle graph \\
Square ($C_4$) & 4 & 4 & Cycle graph \\
Pentagon ($C_5$) & 5 & 5 & Cycle graph \\
Complete $K_3$ & 3 & 3 & Equivalent to $C_3$ \\
Complete $K_4$ & 4 & 6 & Tetrahedron \\
Path $P_4$ & 4 & 3 & Linear chain ($1-2-3-4$) \\
Star $S_4$ & 4 & 3 & Central hub connected to 3 leaves \\
\bottomrule
\end{tabular}
\caption{Graph topologies used for random preference scaling experiments.}
\label{tab:random_topologies}
\end{table}

% Bibliography
\bibliographystyle{plainnat}

\begin{thebibliography}{99}

\bibitem{Condorcet1785}
De Condorcet, N. (2014). \textit{Essai sur l'application de l'analyse \`a la probabilit\'e des d\'ecisions rendues \`a la pluralit\'e des voix}. Cambridge University Press. (Original work published 1785)

\bibitem{Arrow1951}
Arrow, K. J. (1951). \textit{Social Choice and Individual Values}. Wiley.

\bibitem{Baigent1987}
Baigent, N. (1987). Preference proximity and anonymous social choice. \textit{Quarterly Journal of Economics}, 102(1), 161-170.

\bibitem{Chichilnisky1980}
Chichilnisky, G. (1980). Social choice and the topology of spaces of preferences. \textit{Advances in Mathematics}, 37(2), 165-176.

\bibitem{ChichilniskyHeal1983}
Chichilnisky, G., \& Heal, G. (1983). Necessary and sufficient conditions for a resolution of the social choice paradox. \textit{Journal of Economic Theory}, 31(1), 68-87.

\bibitem{Jiang2011}
Jiang, X., Lim, L.-H., Yao, Y., \& Ye, Y. (2011). Statistical ranking and combinatorial Hodge theory. \textit{Mathematical Programming}, 127(1), 203-244.

\bibitem{Bredon1997}
Bredon, G. E. (1997). \textit{Sheaf Theory} (2nd ed.). Springer.

\bibitem{Kashiwara2006}
Kashiwara, M., \& Schapira, P. (2006). \textit{Categories and Sheaves}. Springer.

\bibitem{Rosiak2022}
Rosiak, D. (2022). \textit{Sheaf Theory Through Examples}. MIT Press.

\bibitem{Curry2014}
Curry, J. (2014). Sheaves, cosheaves and applications. PhD thesis, University of Pennsylvania.

\bibitem{Ghrist2014}
Ghrist, R. (2014). \textit{Elementary Applied Topology}. Createspace.

\bibitem{Hansen2020}
Hansen, J., \& Ghrist, R. (2020). Opinion dynamics on discourse sheaves. arXiv:2005.12798.

\bibitem{Weinberger2004}
Weinberger, S. (2004). On the topological social choice model. \textit{Journal of Economic Theory}, 115, 377-384.

\bibitem{Robinson2020}
Robinson, M. (2020). Assignments to sheaves of pseudometric spaces. \textit{Compositionality}, 2, 2. arXiv:1805.08927.

\bibitem{List2002}
List, C. (2002). Two concepts of agreement. \textit{The Good Society}, 11(1), 72-79.

\bibitem{ListPettit2002}
List, C., \& Pettit, P. (2002). Aggregating sets of judgments: An impossibility result. \textit{Economics and Philosophy}, 18(1), 89-110.

\bibitem{ListPuppe2009}
List, C., \& Puppe, C. (2009). Judgment aggregation: A survey. In P. Anand, P. Pattanaik, \& C. Puppe (Eds.), \textit{Handbook of Rational and Social Choice} (pp. 457-482). Oxford University Press.

\bibitem{Fishburn1970}
Fishburn, P. C. (1970). Arrow's impossibility theorem: Concise proof and infinite voters. \textit{Journal of Economic Theory}, 2(1), 103-106.

\bibitem{KirmanSondermann1972}
Kirman, A. P., \& Sondermann, D. (1972). Arrow's theorem, many agents, and invisible dictators. \textit{Journal of Economic Theory}, 5(2), 267-277.

\bibitem{Lauwers1995}
Lauwers, L., \& Van Liedekerke, L. (1995). Ultraproducts and aggregation. \textit{Journal of Mathematical Economics}, 24(3), 217-237.

\bibitem{Mallows1957}
Mallows, C. L. (1957). Non-null ranking models. I. \textit{Biometrika}, 44(1/2), 114-130.

\bibitem{Lu2011}
Lu, T., \& Boutilier, C. (2011). Learning Mallows models with pairwise preferences. \textit{Proceedings of the 28th International Conference on Machine Learning}, 145-152.

\bibitem{Black1948}
Black, D. (1948). On the rationale of group decision-making. \textit{Journal of Political Economy}, 56(1), 23-34.

\bibitem{Bubenik2014}
Bubenik, P., \& Scott, J. A. (2014). Categorification of persistent homology. \textit{Discrete \& Computational Geometry}, 51(3), 600-627.

\bibitem{BrightwellWinkler1991}
Brightwell, G., \& Winkler, P. (1991). Counting linear extensions. \textit{Order}, 8(3), 225-242.
\end{thebibliography}

\end{document}